# Planning Distributed Security Operations Centers in Multi-Cloud Landscapes

## A Case Study

Andreas U. Schmidt[1], Sven Knudsen[2], Tobias Niehoff[2], and Klaus Schwietz[2]

[1] Novalyst IT AG, Robert-Bosch-Str.38, 61184 Karben, Germany
[2] NOZ Digital GmbH, Fördestraße 20, 24944 Flensburg, Germany
`andreas.schmidt@novalyst.de`

**Abstract.** We present a case study on the strategic planning of a security operations center in a typical, modern, mid-size organization. Against the backdrop of the company's multi-cloud strategy a distributed approach envisioning the involvement of external providers is taken. From a security-centric abstraction of the organizational IT-landscape, a novel strategic planning method for security operation centers is developed with an adaptable relationship matrix as core tool. The method is put to a practical test in modeling different levels of engagement of external providers in the center's operation. It is shown that concrete output, such as a core statement of work for an external provider, can easily be derived.

## 1   Background and Introduction

Information security concerns are pressing for any organization. From the management perspective, it is by now common sense to separate recurring and continuously executed IT security tasks in a separate organizational unit, the Security Operations Center (SOC). The SOC shall bundle resources and know-how and reduce organizational complexity in the hope of increased efficiency as well as effectiveness in threat mitigation. Core operational task of the SOC is Security Information and Event Management (SIEM) to continuously monitor IT and swiftly react to attacks. In this paper, we present a case study on the planning of a *distributed* SOC for medium-sized company with considerably complex IT. In accordance with the broad adoption of cloud services by the organization in a multi-cloud strategy, the SOC is, from the outset, envisioned as a distributed entity where internal resources shall be complemented by specialized external service providers. A major planning focus is thus the work distribution between internal and external resources in the SOC. In the course of this planning, we developed a systematic method to structure a distributed SOC, which is generic enough to be applicable to many organizations in a similar situation, but concrete enough to yield tangible results for strategic management.

## 1.1 Task Structure of a Modern Security Operations Center

Following the white paper [1], the history of the "Security Operations Center" goes back to the mid-1970s, to the security requirements of small-scale military information and communication systems. Since then, the technical/organizational functions of an SOC have evolved several decades with an increasing variety of more complex tasks. The SOC has only attracted the interest of applied research for about a decade (see the extensive review article [2] provides a multifaceted overview of the state of knowledge). The practical definition and modeling used here for SOC and the SIEM processes applied in it refers to the basic empirical study [3], wherein the core tasks of an SOC are derived based on empirical surveys of IT practitioners in various organizations and the ENISA guide [4]. In addition, frequently posed practical requirements are collected in [5]. Since the study [3] in 2015, the weighting between tasks has shifted in practice and in the literature (see [2]). In comparison to [3], Figure 1 shows a task structure for an SOC that is partially tailored to the characteristics of the organization examined herein and is therefore more refined regarding the relationship of the SOC to other organizational units (OU). The definition of the (internal or external) OU interacting with the SOC is deliberately kept vague to allow for varying configurations.

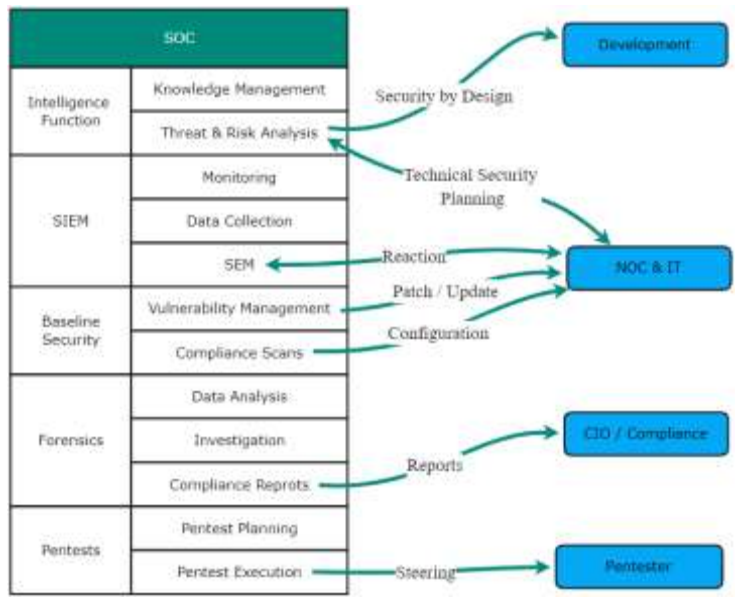

**Fig. 1.** SOC task structure and interaction with OUs.

From this point of view, the SOC has five core functional tasks, which are divided into further sub-tasks:

- The **intelligence function** is a competence center for information security issues in the organization, tasked with the continuous development of security know-how through active knowledge management. The knowledge, comprising both general knowledge of state-of-the-art IT security and specific knowledge of the own IT landscape, is activated by creating attack and risk analyses to advise the Network Operations Center (NOC) and the IT departments in technical and organizational security planning. If the organization carries out own IT developments, this function will promote compliance with security by design principles. The intelligence function can also be used by other OUs, e.g. in the context of security consultations.

- The **SIEM function** is here regarded (differing from [3], where it appeared as a sub-function of monitoring) as a core task of the SOC (cf. [2]). Classic sub-tasks of SIEM are monitoring and systematic data collection (Security Information Management, SIM) and the detection of security incidents with the subsequent (control of) the reaction to them, i.e., mitigation and recovery, (Security Event Management, SEM).
- **Baseline security** integrates a classic task of IT security departments into the SOC. The security of the organization's IT is continuously maintained and improved by monitoring compliance with current security standards and eliminating vulnerabilities in the assigned systems through for example (partially) automated scans. This function interacts with the Network Operations Center (NOC) and IT department for security configuration and updates/patches, allowing for a fluid division of duties.
- The **forensic function** is used for reporting on security incidents and on the security status of the monitored systems. Procedures for data collection, further investigation and finally the securing of digital evidence and reports on this are established by recognized standards [6, 7]. Generated reports are used by the CIO as a basis for decision-making and by the compliance department for mandatory documentation.
- **Pentests** are of paramount importance for securing newly introduced IT systems as well as for safeguarding existing ones. Today, they are often carried out by external actors for two reasons: Firstly, specialized companies have developed a considerable know-how advantage here and secondly, it is now an established practice to separate these tests from the target organization in order to reduce cognitive bias. The SOC plans the type and scope of pen tests and oversees their execution.

Due to their close processual connection to the NOC and IT department, the SIEM and baseline security functions form the operative-technical core of the SOC. The SOC is a complex, labor-intensive work domain combining requirements for technical and organizational know-how and thus for a skillful workforce.

### 1.2  Multi-Cloud Strategy and Changing IT Work Organization

Within the global shift of information technology from organization-internal, on-premise systems to broad usage of cloud services, multi-cloud strategy is arguably seen as a dominating trend for organizations of a certain complexity and size [8, 9]. A multi-cloud strategy selects platforms, infrastructure and applications from different cloud service providers (CSPs), each for a specialized purpose, and integrates them organizationally and technically adapted to the requirements of an organization [10]. The proposed advantages of multi- over single-sourced cloud are manifold [8, p. 6]. Amongst others they comprise reduction of vendor lock-in, cost optimization [11], load balancing and business continuity through partial redundancy, enabling best-of-breed service selection, and even increased security through diversification of data storage and processing [12]. On higher organizational levels it is argued that multi-cloud strategies foster innovation capabilities and agility, e.g., for data analytics [13]. In the following we interpret multi-cloud strategies differently, from the viewpoint of work organization. In our view, this provides the backdrop for the implementation of a modern SOC.

In traditional on-premise IT landscapes, the core task of the IT staff is the deployment, maintenance and monitoring of hardware and software, i.e., classical administration with its frequently recurring work tasks. The creation of interoperability between systems (common and proprietary software) often requires the development of connectors and databases. Core systems often only become usable through special front and back ends, which also have to be developed and maintained in-house. Such necessary work required to be able to use the infrastructure in the first place, is continuous operational expense without added value.

On intermediate stages, hybrid on-premise/cloud system emerge, which outsource (hosted) components in the Platform- and Infrastructure-as-a-Service (PaaS, IaaS) models may also use native cloud applications (Software-as-a-service, SaaS). While the use of PaaS and IaaS only frees IT from a few administration tasks (i.e., the physical computer platform), the software on these platforms must continue to be maintained and administered in the classical way. SaaS components are fundamentally different: they are deployed and maintained through custom *configuration* interfaces. PaaS and SaaS components can often communicate via defined interfaces and standard protocols (e.g., REST), which facilitates the development of connectors.

As a further evolution, the various platforms [14], infrastructures [15], and applications from different CSPs are not only *orchestrated* [16] (i.e., managed and scaled) through unified control interfaces with a high degree of automation, but also *interoperate* seamlessly to enable data processing across clouds (e.g., enabling advanced functionality such as cross-cloud data analytics [13]). This stage is partly visionary at this time[1] and subject of ongoing research efforts [20, 21, 22]. In an ideal multi-cloud environment, IT staff would configure components once at deployment to ensure interoperation and desired functionality, and occasionally for instance at functional changes.

The fundamental change through this evolution is this: Repetitive administration and programming tasks with no added value are replaced by on-demand (at deployment or when functions change) configuration tasks, which immediately enable productivity. Human resources in IT are thus set free for either work that adds value to the organization or in the literal sense. These changes of work environment can change the perspective of people working in IT by opening the view for new possibilities and potential for innovation beyond pure efficiency increases. Obviously, these processes must be actively planned and managed to realize the potential benefits.

---

[1] One functional area in which interoperability is already highly developed in the described sense is Identity and Access Management (IAM). Open standards, in particular OAuth 2.0 [17] and FIDO2 [18], enable the creation of cloud services [19] federating various sources of user identities and credentials to provide multi-factor authentication and so forth.

### 1.3 Contribution, Related Work, and Paper Outline

The multi-cloud strategy sets the backdrop and constraints for the planning of a SOC in our case study. From the outset it is planned to involve external service providers with specialized know-how, technical tools, and capabilities, who take on specific SOC tasks, while the internal SOC remains an important strategic OU. In accordance, internal SOC personnel shall focus on the most critical tasks in the SIEM and baseline security areas of the SOC, and the core strategic tasks in the other areas. The overall aim is to liberate the internal SOC from repetitive operational and bureaucratic tasks to optimize reactions to security events and enable continual security improvements.

Against this backdrop, the present work systematically develops a novel, conceptual planning tool for SOC. Our approach, rooted in systems theory [23], first systematically abstracts the technical elements of the organization's IT by subsequent functional and security-centric grouping into generic function categories or organizational subunits. We then relate organizational subunits to the structure of the SOC (Fig. 1) in a simple relationship matrix. We show the internal consistency of the concept with respect to common security controls applied to the subunits. The tool is abstract and versatile and can be used for different aspects of SOC planning by applying various *organizational dimensions* (e.g., resources, required know-how, etc.) to the matrix cells. We specifically show its application to the qualification of the roles of internal and external SOC actors, respectively, i.e., for planning the distributed SOC described above. Finally, as a concrete use, we show how a statement of work for an external SOC service provider can be derived from the planning matrix by systematic reduction of abstraction.

Complementary to our work, the study [24] presents a scheme to empirically characterize existing SOCs according to seven defined organizational dimensions, but independently of the organizational IT background. At the end of the SOC planning process decision making regarding, e.g., the selection of service providers begins. A typical, evolved scheme for this, which can complement our planning method can be found in [25], or see [26, 27] for quantitative decision-making methods. In contrast to granular, formal approaches to process planning for IT security systems [28], our approach is strategically informal (but could serve as a starting point for formalization. Conceptually, we follow the broad lines of system-theoretic modeling for IT governance, cf. [29] with a hands-on approach.

We start out in section 2 with a high-level description of the IT landscape in the present case study, separating the internal systems from the various CSP domains employed. A more detailed description of system elements and their placement into groups of common functions follows, which yields the set of IT organizational subunits as a basis for the subsequent conceptualizing. Section 3 presents the core method for SOC planning, starting with a generic reduction of complexity by further abstraction of function groups into broader function categories of common security characteristics. The desired relationship matrix is then constructed, and its consistency with respect to those security characteristics is rationalized. In section 4, the matrix is concretely applied as a planning tool for a distributed SOC. Section 5 concludes with a discussion of limitations and shortcomings of the presented concepts, some suggestions for further extensions, potential applications, and further research.

## 2    The Case at Hand

Our case study concerns a medium-size newspaper publishing house with about 3000 IT end-users distributed over two main and a few subsidiary sites, also working mobile or in home-office. The technical facilities of the organization range from in-house print production[2], digital content management, administrative and office applications, to a state-of-the-art digital online publishing system, and thus form a rather complex, heterogeneous technical landscape. We believe that this status quo is rather typical for many medium-sized organizations at this time. Four major sectors can be identified:

The **internal IT** systems span across all technical and application layers from production machinery to content production, administrative, and other business software and is managed and maintained in a common network spanning various physical locations. **Google Workspace** (GWS) has been rolled out company-wide as a universal tool for day-to-day office work and communication. On infrastructure and technical layers, GWS is increasingly used for user identity management and the management of end-user devices. The third zone includes **special cloud applications** and services at different levels. Parts of operational business, for example in marketing, have been outsourced to special cloud services. As an important infrastructure part, a cloud service for Identity & Access Management (IAM) provides seamless and secure user logon and authentication across the different sources of identities and credentials used within the organization (e.g., Google and Active Directory). Additionally, some special cloud-based tools for IT Security are used, e.g., for endpoint security monitoring. Finally, **Amazon Web Services** (AWS) as a versatile cloud resource is the foundation of the company's Digital Publishing Platform (DPP). This complex core system of digital production consists of a number of connected components (loosely coupled in a service-oriented architecture [29]), partly own developed, partly contributed, and operated by external partners. The AWS infrastructure itself is managed by a specialized firm. In the development of the DPP, the change of work organization described in Sec. 1.2 has been most widely realized, exemplary with the own implementation of the online sales and subscriber management system – a core techno-economic asset of the DPP. Here, own application logic for operational business has been developed and integrated in cloud-native applications, employing modern methods such as serverless computing [31] without any consideration of the underlying infrastructure.

The four zones may be further broken down into functional groups, while maintaining a rough orientation on the layer model of information technology (although function groups often contain technical elements across different layers). There are many other possibilities to structure the IT landscape, e.g., technical-architectural (network segments and zones), or security-related such as access control groups and hierarchies. The functional grouping chosen here serves as a basis for strategizing the SIEM processes and SOC organization and is driven by two boundary conditions of the overarching security strategy: Functions are to be grouped and classified i) w.r.t. their criticality for business continuity, and ii) so as to enable group-wise reporting on security posture and security events for compliance purposes. These external conditions must be compatible with the conditions for a well-functioning SOC: iii) The function groups allow for coherent security operations regarding the core SIEM processes, i.e. they allow for application of common security controls and, for instance, technical expert knowledge to mitigate security events (we will check on this condition later in this section). With these conditions in mind, we identify twelve function groups.

---

[2] Which is strategically *not* fully outsourced due to the regional locality of some news outlets in the company's portfolio.

1. Print **production machines** are special elements at the network edge: Control data goes in, telemetry data goes out, and network connection is almost exclusively via special control servers. Exposure units, stamping and bending machines and the actual printing machines are connected via various interfaces, e.g., an SCSI interface connected to a dedicated computer via a SCSI-USB converter. Printing presses often also have Ethernet interfaces and can speak TCP/IP. Configuration and maintenance is carried out via control servers or physically on site.
2. **Print production systems** form the IT environment of print production machinery and thus the network perimeter of the group 1 described above. The group contains highly specialized applications such as integrated workflow control as well as prepress/exposure/quality control software. In addition, there are infrastructure components such as font servers (type servers) and (Windows) file servers for data exchange with external providers.
3. The **telephony** group combines the various Voice-over-IP (VoIP) infrastructures and applications throughout the company, among them infrastructure elements such as a number of SIP servers (telco system software, PBX) and session border controllers from different manufacturers. A unified communications solution (telephone switchboard) is used at the application infrastructure level. Call center solutions form the application layer of this group. This group is a vertically integrated stack through all network layers from hardware to application and is largely isolated from other components. Relations to other functions exist at infrastructure and application levels (e.g., with directory services and CRM systems). This structure renders this group a candidate for outsourcing and/or virtualization.
4. **Network Infrastructure** comprises basic network infrastructure functions at TCP/IP and higher protocol levels. These include routers, switches, DHCP, and DNS servers. The established logical network structures, subnets and associated network security areas (Demilitarized Zones, DMZ) are also located here. This function group cannot be fully outsourced. Finally, it also includes the platforms on which higher-level services and applications run, i.e., the (few) physical servers and the large number of own-managed virtual machines (VMs).
5. A **remote desktop service** provides device- and location-independent virtual Windows and Linux desktop environments for devices outside the internal network, which only have access to internal resources via this secure service. Remote desktop agents on end devices provide a secure environment for the execution of applications delivered by the service. The remote desktop service infrastructure is heterogeneous, consisting of internal servers and cloud services of the provider[3].

---

[3] The service provider is Citrix. For an architecture overview, see [32].

6. **End devices** permanently installed in the company network are managed with automated scans, monitoring, and software distribution solutions. Mobile clients – laptops, tablets and other devices with various operation systems – are assigned to unique owners and managed via the device management functions of GWS. Access to internal resources from these clients is secured via remote desktop or VPN.
7. To set up ad hoc **test environments**, personnel of some OEs is privy to set up and administer virtual machines on certain servers. Being used for tests of third-party software and for own development, test systems may interact with internal systems or non-productive replicas thereof. During test and development in the field of print production, occasionally, test systems have to communicate with printing machines in the final stage. Outsourcing of test environments to a cloud-based test farm is a possibility that is currently considered for security and cost-efficiency reasons.
8. **IT Security**, as an OE, governs a variety of security and security management tools ranging from multi-factor authentication solutions and user credential management over multi-platform endpoint security monitoring, virus scanners, patch management and software distribution to passive and active baseline security functions such as firewalls and VPNs. Elements of this group are access-restricted to privy personnel of the OE and isolated within the organization's intranet.
9. **Data Security** is implemented on two levels: While a backup solution regularly secures large amounts of business relevant data, another solution for continuous data protection (CDP) also secures machine images for disaster recovery[4]. The two complementary systems are architecturally similar: Target data sources are connected via proxy servers while management of the sources and storage load balancing is performed by orchestration servers. Short term security in-house storage is complemented by Long-Term Repositories (LTRs), which mostly reside off-shore.
10. The **application infrastructure** contains classic common functions such as mail server (MS Exchange), directory services (Active Directory, AD), print server, Windows domain controllers, database servers and specialties such as local license management servers for Windows (e.g. Office) software. Internal AD is complemented by cloud-based, federated IAM [34, 19] for cross-service access. FTP servers are maintained at various locations, e.g., for forwarding print data to production. Although normally considered an element of the higher application layer, we include GWS in this group, since its functions are commonly used throughout all OEs, and it is managed similarly to the aforementioned infrastructures.
11. The purpose of **static data management** is the secure long-term storage of important documents (incoming and outgoing invoices, personal files, and contracts) from the operative business for compliance. A data management solution for these documents also enables workflows for deadline-based processing. Another solution provides digitalization of paper documents. Both are linked to each other and to higher level business applications, in particular SAP. Further special solutions are employed for secure long-term archiving of documents and email communication.

---

[4] The backup solution is VEEAM; CDP is implemented with Zerto. For a (however biased, marketing-oriented) comparison, see [33].

12. **Business applications** denote the IT functions for value creation proper on the one hand and for daily administration on the other. For the former, a special application for digital asset management (DAM) is the central element enabling cross-media management and processing of content and for newspaper editors and journalists to create it originally. The in-house part of the DAM is used for management, storage and rights control, while another part outplaced to the AWS cloud is tightly integrated with the DPP mentioned above and serves as a content repository for it but also as a frontend for content creators. Data is held redundantly in and replicated regularly between the two parts. The DAM and DPP are accompanied by auxiliary functions – some in-house some outsourced – for advertisement management, digital marketing and campaigning, and logistics planning for printed media distribution. The heart of administration functions is an SAP system with many subfunctions including financial accounting, cash register system, material planning, wages/salaries, controlling, and business intelligence. Organizationally, SAP is – as per common practice – operated and administered by a separate in-house team. SAP connect externally via dedicated DMZ routers. Apart from SAP, a separate application is used for human resources management.

The empirical part of the case study concludes with an evaluation of the functional groups from the perspective of security operations. Particularly, we want to understand how relevant a functional group is for the SOC – in terms of resources needed and effect exerted, and which particular methods, i.e., security controls [35], are best applied to each group. For the latter, we identify seven security controls that are tasks of the SOC within the functions of SIEM and baseline security. They are (with mnemonic tokens, the first four from the SIEM function, the last three from baseline security)

- **S.Com**: Communication Monitoring relies on intercepting and examining data communications between network nodes and examining it for anomalies. Many levels apply, from rough logging of data volumes and network loads to examining individual IP packets with Deep Packet Inspection (DPI).
- **S.Acc**: Access Monitoring observes the triggering of functions on a node by requests from another node or a user. This necessitates inspection of communication protocols (e.g., DNS, LDAP, SMTP, SMNP, HTTP, SOAP, REST, and application-specific protocols) at all layers. Target systems mostly carry logs that are data sources for SIM. IAM systems also provide granular logs. Checking plausibility and detecting anomalies of access attempts is a core element of intrusion detection (see below).
- **S.App**: Application Monitoring yields information about security-related events in business applications. Extensive logging of application processes plays the main role here. Occasionally, applications also have their own security modules, which, e.g., "scan" the application's configuration or databases. For SIM, the supplied logs (in proprietary formats) must be individually accessed which can be cumbersome.
- **S.ID**: Intrusion Detection is a meta-method that combines data from the three aforementioned methods and additional data obtained from endpoints (endpoint security) in order to detect breaches of system security by attackers. This is a part process of SIEM which is often effected by separate intrusion detection systems (IDS).

10- **B.Ept**: Endpoint Security comprises a variety of tasks such as setting up (restricting) user access rights, configuring system and software regarding to security features (e.g., virus scanners), and monitoring the clients during operation. Special products for remote monitoring and management of endpoints install agent programs on endpoints. Such agents can often trigger alerts or provide logs for SIEM purposes.
- **B.Vuln**: Vulnerability Management refers to the detection and elimination of security deficiencies – entry points for attackers – through updates or reconfiguration. Information about vulnerabilities mostly comes from external (general/public [37] or manufacturer-specific) sources that have to be constantly monitored. Vulnerabilities are security-relevant events and their management is part of SIEM.
- **B.Peri**: Perimeter Security essentially consists of two parts. The static part restricts the communication between nodes in the network, for example by setting up separate segments or only allowing protected (VPN) connections. The dynamic part secures perimeters by detecting and filtering unauthorized communication, e.g., by firewalls. Both parts provide relevant information (network access and firewall logs) for SIM.

With these definitions, three questions were put to an internal expert committee[5]:

— What is the primary security control which *must* be applied to a function group?
— What are (at most two) secondary controls which *should*[6] be applied to a group?
— What is a group's *relevance* to security operations, ranked as *low*, *medium*, or *high*?

The last question asks for some elaboration. The committee was here asked for a synthetic opinion (euphemism for "gut feel") on the effort incurred by security operations on a function group. Three subsidiary valuation criteria were provided: i) Criticality of the function according to common confidentiality-integrity-availability schematics, ii) applicability and effectiveness of the seven security controls, and iii) expected workload in terms of attack frequency times mitigation effort. For each of the questions, the committee was asked to provide a rationale for its answer, a requirement which serves as a guidance toward rational group consensus. The result of the survey is shown in Table 1. The Abstract Category (Cat.) column is introduced in the following section.

**Table 1.** Expert survey on applicable security controls, and relevance to security operations.

| Cat. | Function Group | Prim. Ctrl. | Sec. Ctrl. | SO Relevance |
|------|----------------|-------------|------------|--------------|
| A | Production machines | B.Peri | B.Vuln | Medium |
| E | Print production | S.ID | B.Vuln | High |
| A | Telephony | B.Peri | S.ID | Medium |
| B | Network Infrastructure | B.Ept | B.Vuln, S.ID | High |
| D | Remote Desktop Service | S.Acc | B.Vuln, S.ID | Medium |

---

[5] Consisting of members from the IT security, IT proper, and, if applicable, the specialty department responsible for use and/or management of a function group. Proportionality was disregarded since the group was asked to respond with rational consensus.

[6] The terms *must* and *should* are used here in accordance with the IETF definitions of normative language [36] which was familiar to most participants, or otherwise introduced to them.

| | | | | |
|---|---|---|---|---|
| B | End Devices | B.Ept | B.Vuln, S.ID | High |
| C | Test Environments | B.Peri | S.Acc, S.Com | Low |
| C | IT Security | B.Peri | S.Acc | High |
| C | Data Security | S.Com | S.App, B.Vuln | High |
| D | Application Infrastructure | S.Acc | B.Vuln, S.ID | High |
| D | Static Data Management | S.Acc | S.App, B.Vuln | Low |
| E | Business Applications | B.Vuln | S.ID | High |

## 3   A Strategic SOC Planning Method

The concrete project target, based on the empirical findings above, is to plan explicit requirements for the involvement of external (multiple) providers in the SOC, and the work distribution between the providers and the internal SOC. For this, we create a tool to structure and inform strategic discussions for SOC planning. The idea is very simple: SOC tasks (see Sec. 1.1) are orthogonally related to the identified function groups. The resulting abstract relationship matrix (a tool commonly used in operations research, see, e.g., [38]) can then be filled with various conceptual dimensions, e.g., task priority, criticality, resource requirements, etc. The populated matrices can then support strategic decisions regarding SOC tasks in the context of the concrete IT landscape. As an abstract method, this approach can be used in many steps of the IT management process, from organizational and resource planning to downstream evaluation. However, human strategic decision-making, particularly in group discussions on the executive level, can hardly be successful with a relationship matrix representing the SOC tasks and IT functions in a 12x12 matrix. Thus, we take a further abstraction step and relate the five SOC task groups to five abstract function categories, reducing the function groups. The categories are assembled according to similarity, in turn measured by counting the common elements in the sets of all entries in the rows of two compared function groups. For five target categories, this leads to the following abstraction:

A. Operational Technology (OT): Production Machines, Telephony[7]
B. Common Infrastructure (Infra): Network Infrastructure, Clients
C. Core Security (Sec): IT Security, Data Security, and Test Environments
D. Application Services (Serv): Application Infrastructure, Remote Desktop Services, and Static Data Management
E. Business Functions (BF): Business Applications, Print Production

This high-level abstraction has – apart from facilitating decision-making – the potential advantage of making the SOC relationship matrix applicable to a wider context of organizations than the particular one considered herein. However, the possibility of critical information loss should be carefully considered with such a broad abstraction. Errors may manifest themselves essentially in two ways. Firstly, information loss may render the function categories homogeneous w.r.t. SOC tasks. Thus, essential – technical and/or organizational – properties of IT functions that differ within a category may be abstracted away. As a result, tasks assigned to the SOC in a cell may for example require differently skilled personnel for the different functions in the category. This may lead to resource misallocation especially within the SIEM and baseline security tasks (which are used for category-building). Secondly, the concepts on which the category abstraction is based as well as the definition of the tasks of the SOC, are subject to changes due to technical and scientific progress or changing economic boundary conditions. Here, we address the first concern in two ways. In this section, we carry out a thought experiment corroborating the consistency of the SOC relationship matrix with the empirical data of Table 1. In the next section we show that the information reduction incurred in the abstraction is, in principle, reversible.

---

[7] OT is commonly understood as such hard- and software that detects or causes changes in industrial production systems. In the present case, telephony is a frequently used as one communication means on and to the production floor both for maintenance and production steering. Hence, we count telephony into OT here.

**Table 2.** SOC Relationship Matrix

| Organizational Dimension | | SOC Main Task | | | | |
|---|---|---|---|---|---|---|
| | | Intelligence | SIEM | Baseline Security | Forensics | Pentests |
| Category | A OT | | 0 | 6 | | |
| | B Infra | | 3 | 9 | | |
| | C Sec | | 4.7 | 3.7 | | |
| | D Serv | | 6 | 2 | | |
| | E BF | | 4.5 | 4.5 | | |

Table 2 presents the SOC relationship matrix proper with numerical entries representing a simplistic score that may coarsely be interpreted as expected effort required by the respective SOC task for a category. It is based on the assignments of security controls to the SIEM and baseline security task in Table 1 respectively, and the overall SO relevance of the function groups in a category. The detailed scoring method is rather immaterial[8], and the results should not be over-interpreted. The exercise shows, however, that the function categories are clearly *discernible* from each other, even with this extremely lossy evaluation.

## 4  Application: Planning Distributed SOC

In the following, the SOC relationship matrix is used specifically for modeling the possible division of tasks between an internal SOC and external service providers. The organizational dimensions in the matrix cells should accordingly reflect this desired division of labor. For this, the SOC sub-tasks (on the level below the five main tasks) in each cell are qualitatively divided into four contribution levels:

- I: The task is mainly carried out by internal staff;
- IE: The task is mainly carried out/led by internal staff, externals support/collaborate;
- EI: The task is mainly carried out/led by externals, internals support/collaborate;
- E: The task is mainly carried out by external parties.

---

[8] For each function group, a secondary control entry yields 1 point, a primary one 2 points for the associated SOC task. The sum is multiplied by the SOC relevance of the group with the factors 1, 2, and 3 for "Low," "Medium," and "High," respectively. The result is normalized regarding the number of function groups in the respective category, i.e., divided by 3 or 2.

A middle level of equal division of labor is deliberately omitted here in order to avoid the problem of diffusion of responsibility in organizational decision-making. According to the task division each cell is assigned a numerical value representing the overall involvement of the external providers, depending on the aggregate involvement levels of subtasks, and according to a simple, 5-level evaluation scale: 0.1: marginal, 0.3: low, 0.5: equivalent, 0.7: predominant, 0.9: central. With these preparations, planning workshops were held with a senior staff group of responsible experts from IT and IT security departments. They developed three different models for the participation of external service providers in the SOC[9]. The method of rational consensus was employed again, and the group was asked to provide brief rationales for their decisions.

**Model 1 "status quo."** For the first model, participants were asked to model the *status quo* within the framework (mainly to get acquainted to the method). Table 3 shows the result first model evaluation. The rationales for the valuations are given per SOC main task group below the table.

**Table 3.** "Status Quo" Model. The abbreviations for the SOC subtasks used are as follows: Intelligence Function: Knowledge Management, KM; Risk Analysis, RA. SIEM: Monitoring, Mo; Data Collection, DC; SEM, S. Baseline Security: Vulnerability Management, Vu; Compliance Scans, CS. Forensics: Data Analysis, DA; Investigation, In; Compliance Reports, CR. Pentests: Planning, Pl; Execution, Ex. The aggregate numerical valuation is abbreviated as V and is also indicated by shading of the cells.

| External Provider Involvement | | SOC Main Task | | | | |
|---|---|---|---|---|---|---|
| | | Intelligence | SIEM | Baseline Security | Forensics | Pentests |
| Abbreviation | | V KM/RA | V Mo/Da/S | V Vu/CS | V DA/In/CR | V Pl/Ex |
| Category | A OT | 0,1 I/I | 0,1 I/I/I | 0,1 I/I | 0,3 IE/IE/IE | N/A[10] |
| | B Infra | 0,3 IE/I | 0,3 IE/IE/I | 0,3 I/IE | 0,3 IE/IE/IE | 0,7 IE/E |
| | C Sec | 0,1 I/I | 0,1 IE/I/I | 0,3 I/IE | 0,3 IE/IE/IE | 0,7 IE/E |
| | D Serv | 0,3 IE/I | 0,3 IE/IE/I | 0,3 I/IE | 0,3 IE/IE/IE | 0,7 IE/E |
| | E BF | 0,3 IE/I | 0,1 I/I/I | 0,1 I/I | 0,3 IE/IE/IE | 0,7 IE/E |

---

[9] It should be noted that the external entities are always thought of in the plural, although this is not explicit in the models. This allows for further model differentiation including the fact that already employed providers play roles in the SOC, for example as suppliers of know-how (e.g. security bulletins [39], with which GWS contributes to knowledge management) or data (logs from AWS that go into SIEM data collection).

[10] The group agreed that pentests in this category are to be excluded from the planning exercise since it would require physical access to installations or (for telephony) making phishing calls. It was decided to restrict planning to pentests carried out over digital channels, exclusively.

- Intelligence: External parties are only involved as sources of information to the extent that they already provide up-to-date information about security features during normal operation, e.g., as part of the ongoing management of the cloud infrastructure anyway, for example through regular security bulletins by a cloud provider. Internal IT Security department manages core security independently.
- SIEM: External services are used at most for data collection and monitoring and only where this is possible due to the general accessibility of the systems without additional effort (e.g. automated logs and reports from AWS and GWS). The SEM capability is operated managed internally using specific internal security tools.
- Baseline Security: The OT group and Business applications can only be protected by special internal know-how. For infrastructure, core security and application services, there are external tools that provide support, particularly with scans.
- For forensics, an external service provider will only be used on a case-by-case basis if an external, independent investigation of an incident is required.
- Pentests are only planned internally with precise requirements, but – based on the current IT security methodology – are carried out by external parties.

**Model 2 "maximum external involvement."** Here, participants were asked to consider the largest currently technically possible (maximal) involvement of external providers. In the model shown in Table 4, essential and core tasks are outsourced to the external provider.

Table 4. Maximum external involvement model.

| External Provider Involvement | | SOC Main Task | | | | |
|---|---|---|---|---|---|---|
| | | Intelligence | SIEM | Baseline Security | Forensics | Pentests |
| Abbreviation | | V KM/RA | V Mo/Da/S | V Vu/CS | V DA/In/CR | V Pl/Ex |
| Category | A OT | 0,3 IE/IE | 0,3 I/IE/IE | 0,3 IE/I | 0,5 IE/EI/EI | N/A |
| | B Infra | 0,9 E/EI | 0,9 E/E/EI | 0,7 E/EI | 0,7 E/EI/EI | 0,9 EI/E |
| | C Sec | 0,7 EI/EI | 0,9 E/E/EI | 0,7 E/EI | 0,7 E/EI/EI | 0,9 EI/E |
| | D Serv | 0,9 E/EI | 0,9 E/E/EI | 0,7 E/EI | 0,7 E/EI/EI | 0,9 EI/E |
| | E BF | 0,7 EI/EI | 0,7 EI/E/EI | 0,5 EI/IE | 0,7 E/EI/EI | 0,9 EI/E |

The organization becomes largely dependent on external service providers in this model, as can be seen in the evaluation rationales:

- Intelligence: In the areas of predominant competence (infrastructure, core security, and application services), the service provider plays the leading role and may report directly to management. The internal SOC only plays a more than marginal role where special access to internal functions applications is necessary.
- The external SIEM provider has tools allowing autonomous collection of data and monitoring of all core elements of the IT landscape, including the essential cloud services. Additionally, the provider plays a leading role in the SEM by alerting and also specifying (and as far as possible implementing) the reactions to security incidents. The role of the external provider is only limited by access to internal systems.

- Baseline Security: The service provider is able to carry out compliance scans with own tools and has competence on current vulnerabilities of all core systems. For fixing (patching) vulnerabilities, the internal SOC may still play a minor role.
- In forensics, an external service provider regularly has a leading role, conducting all investigations independently, with the internal SOC only acting as the client (except for the OT world for which internal staff may need to assist with data acquisition).
- For pentesting, service providers are leading both planning and implementation. They extensively use their knowledge of the organization's IT from their involvement in the other SOC tasks (esp. SIEM and baseline security). To carry out independent pentests, a planning provider would, if necessary, commission another actor. The internal SOC accepts test results and initiates corresponding activities.

**Model 3 "plan target."** The third model should represent a "middle ground," as a basis for planning of service providers' contributions to the SOC in a short to medium timeframe. I.e., the third model created by the group is the actual planning target. The result is shown in Table 5.

Table 5. Medium involvement of external providers.

| External Provider Involvement | | SOC Main Task | | | | |
|---|---|---|---|---|---|---|
| | | Intelligence | SIEM | Baseline Security | Forensics | Pentests |
| Abbreviation | | V KM/RA | V Mo/Da/S | V Vu/CS | V DA/In/CR | V Pl/Ex |
| Category | A OT | 0,1 I/I | 0,1 I/I/I | 0,1 I/I | 0,3 IE/IE/IE | N/A |
| | B Infra | 0,5 EI/IE | 0,7 EI/EI/IE | 0,5 IE/EI | 0,7 EI/IE/EI | 0,7 IE/E |
| | C Sec | 0,3 IE/IE | 0,5 IE/EI/IE | 0,3 IE/IE | 0,5 IE/IE/EI | 0,7 IE/E |
| | D Serv | 0,5 EI/IE | 0,7 EI/EI/IE | 0,5 IE/EI | 0,7 EI/IE/EI | 0,7 IE/E |
| | E BF | 0,3 IE/IE | 0,5 IE/EI/IE | 0,3 IE/IE | 0,5 IE/IE/EI | 0,7 IE/E |

- Intelligence function: Extended competence of the service providers in the central areas of infrastructure and application services is used. A suitably qualified service provider can contribute to the strategic SOC tasks through special risk analyses.
- SIEM: Service providers have specific strengths in the infrastructures and application services areas, which they monitor with their own technical means and take on the core tasks of SIM. Dedicated internal staff takes on roles responding to security incidents (SEM), e.g., with response plans, guided by the provider. Service providers can also and specifically monitor the cloud services (in Business functions).
- The service providers are able, through their extended technical means, to carry out security scans of the monitored systems regularly or on request and to report on them. In addition, they independently collect information that contributes to vulnerability management. Internal SOC is still responsible for patches and updates.
- The service provider's technical capabilities enable it to contribute significantly to forensic data analysis and compliance reporting in the areas of its core competencies. A provider is tasked with collating reports on a case-by-case basis.
- With regard to pentests, there is no significant change compared to model 1. The SOC service provider may contribute to the planning in the areas of his highest competence, but this does not change the overall assessment.

As a concrete use of the abstract model planning, we derive a statement of work (SoW) from model 3. Methodologically this shows that the information reduction involved in the various abstraction levels on which the SOC relationship matrix rests is in principle reversible. For the IT function groups this is simply done by making the system description in section 2 and the category definitions a basis of the SoW. For the SOC tasks, detailed definitions of subtasks yield descriptions of providers' duties. For SIEM and baseline security for instance, the seven security controls of section 2 can be used for this description. Others are found in the relevant literature [2, 3, 4, 5]. These details are weighted with the contribution levels for the subtasks to obtain a SoW task description for each cell in the planning matrix. As an example, the common SoW task description for common infrastructure and application services for SIEM is as follows:

"For the systems in the defined categories Common Infrastructure and Application Services, the contractor performs the following tasks:

1. The contractor independently monitors the systems via all accessible (standardized or proprietary) interfaces and uses its own technology for the monitoring. The monitoring includes at least i) the operational readiness of the monitored systems, ii) data traffic, and iii) access to their resources (from internal and external, wherever possible and applicable).
2. The client enables data and/or physical access to all of the above systems (protocols, ports, access authorizations) necessary for the fulfillment of the tasks mentioned.
3. The contractor collects all accessible and security-relevant data from the above systems, such as logs, status messages (to be actively queried as part of the monitoring) or error reports. He structures the collected data and makes it available to the client in digital form if required.
4. The contractor operates a state-of-the-art SEM system to identify current security threats and possible attacks on monitored systems. In the event of a recognized security incident, the contractor immediately provides the client with an incident report in a standardized form containing at least the following information: i) Type of incident, ii) Affected systems, iii) Criticality/Risk Assessment, iv) Allowable reaction time v) Available information about the attacker.
5. The contractor supports the client in mitigating identified security incidents. For this, he proposes methods and step-by-step measures in a form that is easy to understand and implement for the client's staff. Highly available and responsive support by the contractor is desired."

## 5     Conclusion

With the results presented herein – a medium-term goal for the involvement of external services and concrete formulations of division of labor – the first part project of the overall project to set up a distributed SOC is complete. The relationship matrix has proven to be an effective tool, and we plan to use it further in the process for provider selection (e.g. to highlight strength/weaknesses of providers in different areas), detailed task formulation, resource allocation, progress and target control, and so on. The structured information gained from our plan process can for instance be a suitable basis for decision-making and provider selection procedures such as in [25].

As already mentioned, the coarseness of the abstractions may lead to errors. Capabilities of service providers could be over- or underestimated and internal resources could be misallocated. We attempt to mitigate adverse effects by continuously involving in quality control of the relationship matrix, especially to ensure best possible fidelity of the abstract picture. This also ensures that the matrix remains up to date regarding the current status and developments of the organization's IT landscape.

As a system-theoretic planning approach, the abstract method developed here seems applicable in more general contexts beyond IT security – broadly whenever the task structure of the OU which *services* a *target subsystem* of the organization is well-defined, and the target subsystem's structure allows for coherent abstraction. For instance, similar planning of a distributed NOC would be natural, but more far-reaching extensions such as looking at the relationship of the human resources department to the target subsystem "personnel" could be envisioned. Concludingly, a remark on a principle trait and limitation of the method: System theory as applied here is inherently *conservative* in that it is unable to transcend the boundaries of the concepts in which the system and its elements are defined. One cannot plan for revolutionary changes with this method.